\documentclass[%
 reprint,
 amsmath,amssymb,
 aps,
 prl,
floatfix
]{revtex4-2}

\usepackage[T1]{fontenc}
\usepackage{graphicx}
\usepackage{dcolumn}
\usepackage{bm}
\usepackage{hyperref}

\usepackage{soul,xcolor}

\allowdisplaybreaks
\DeclareMathOperator{\sinc}{sinc}

\usepackage{verbatim}

\begin{document}


\title{Resolving the Orientations of and Angular Separation between a Pair of Dipole Emitters}

\author{Yiyang Chen}
\author{Yuanxin Qiu}%
\author{Matthew D. Lew}
\email{mdlew@wustl.edu}
\affiliation{%
 Preston M.\ Green Department of Electrical and Systems Engineering, Washington University in St.\ Louis, Missouri 63130, USA
}%

\date{\today}

\begin{abstract}
We prove that it is impossible to distinguish two spatially coinciding fluorescent molecules from a single rotating molecule using polarization-sensitive imaging, even if one modulates the polarization of the illumination or the detection dipole-spread function (DSF). If the target is known to be a dipole pair, existing imaging methods perform poorly for measuring their angular separation. We propose simultaneously modulating the excitation polarization and DSF, which demonstrates robust discrimination between dipole pairs versus single molecules. Our method improves the precision of measuring centroid orientation by 50\% and angular separation by 2- to 4-fold over existing techniques.
\end{abstract}

\maketitle



Single-molecule (SM) nanoscopy has become invaluable for overcoming the optical diffraction limit and observing nanoscale structures and dynamics within biological systems \cite{mockl_2020jacs_super,lelek_2021nrmp_single}. Leveraging the resolvability of blinking molecules over time, these methods repeatedly detect and localize isolated point-spread functions to determine molecular positions \cite{steves_2024arpc_single-molecule} and/or orientations \cite{brasselet_2023optica_polarization,wu_2024_dipole-spread,zhang_single-molecule_2024}. However, leveraging molecular blinking necessarily sacrifices temporal resolution \cite{sage_2019nm_super-resolution} for improved spatial resolution, thereby limiting the ability to observe dynamic processes in biological systems \cite{mockl_2022boe_deep,saguy_2023nm_dblink}. Resolving emitters beyond the Abb\'{e} diffraction limit remains an active research area \cite{jusuf_2022oe_towards, goldenberg_2022ic_learning}. 

Recent studies have established fundamental quantum estimation limits for resolving two incoherent point sources in 2D \cite{tsang_2016prx_quantum, ang_quantum_2017, lupo_2016prl_ultimate} and 3D \cite{yu_2018prl_quantum, zhou_2019optica_quantum-limited, liang_2023oe_quantum}, and several demonstrations have exhibited resolution performance approaching theoretical limits \cite{nair_2016oe_interferometric, tham_2017prl_beating, rouviere_2024optica_ultra}. 
As MINFLUX and related techniques \cite{balzarotti_2017science_nanometer, gwosch_2020nm_minflux, weber_2021np_minsted, weber_2023nbt_minsted} approach ångström-level resolution, here, we explore an orthogonal approach: how well can the angular separation between two spatially coinciding dipole emitters be resolved using polarized light? Modulating the polarization of an illumination beam has been demonstrated to improve spatial resolution \cite{hafi_2014nm_fluorescence, frahm_2016nm_polarization, hafi_2016nm_Reply}, but to our knowledge, the physical limits of resolving a pair of dipole emitters based upon their separation in orientation space have yet to be established. Given numerous recent developments in measuring dipole orientations \cite{zhanghao_2016lsa_super-resolution,curcio_2020nc_birefringent, ding_2021jpcb_single-molecule, hulleman_2021nc_simultaneous, guan_2022lsa_polarization, wu_2022optica_dipole-spread-function, thorsen_2022boe_photon, zhang_2023np_six-dimensional, jouchet_2023oc_combining, zhong_2024photonix_three-dimensional}, such a theory would provide tremendous guidance for improving measurements of single-molecule rotational dynamics at the nanoscale.

In this work, we derive a simple mathematical proof showing that any technique that solely 1) modulates the phase and/or polarization of fluorescence emission, called dipole-spread function (DSF) engineering, or 2) modulates the polarization of pumping light, termed excitation modulation (ExM), cannot distinguish between a pair of dipole emitters versus a single rotating dipole. This degeneracy stems from the incoherent detection of fluorescence photons and cannot be overcome without prior knowledge of the target sample. Further, even if one is certain that an image originates from a pair of dipoles, our analysis shows that no existing technique can measure their separation angle with high precision. To address these issues, we propose combining ExM and DSF engineering to both distinguish a single wobbling molecule from a pair of molecules and significantly improve the precision of measuring the centroid orientation of and angular separation between a pair of molecules. 

\begin{figure}[tbh]
    \centering
    \includegraphics[width=2.8in]{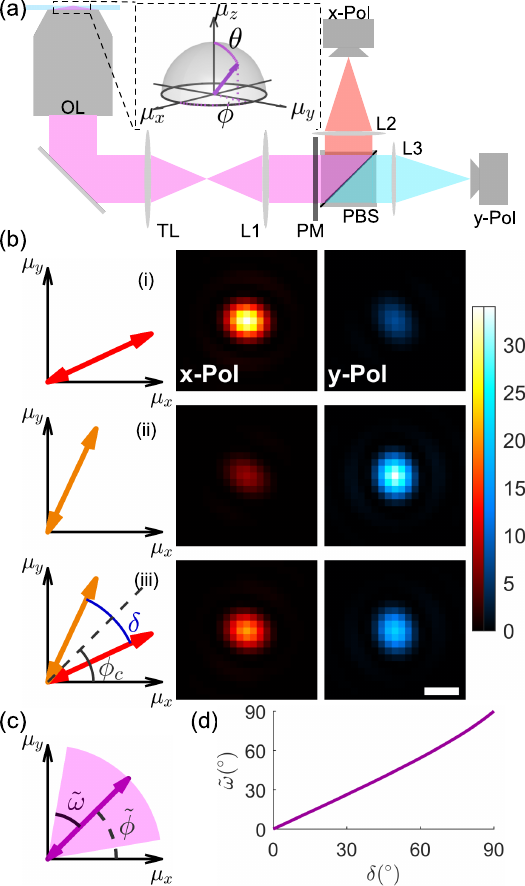}
    \caption{(a)~A polarization-sensitive fluorescence microscope. Fluorescence photons emitted by a dipole are collected by an objective lens (OL) and focused by a tube lens (TL) to an intermediate image plane. A 4f system (L1-L3) with a phase mask (PM) placed at the conjugate back focal plane modulates the collected photons, and a polarizing beam splitter (PBS) separates the photons into (red)~x- and (blue)~y-polarized imaging channels collected by separate cameras. Inset: Orientation coordinates of the transition dipole $\bm{\mu} = \left( \sin{\theta}\cos{\phi}, \sin{\theta}\sin{\phi}, \cos{\theta} \right)$. (b)~(Left)~Schematic and (right)~clear-aperture images of (i)~dipole 1, (ii)~dipole 2, and (iii)~a pair of dipoles separated by angle $\delta$ in the $\mu_x$-$\mu_y$ plane with 500 total detected photons in all cases. Scale bar: 300~nm. (c)~The dipole pair in (b)(iii) is equivalent to a single dipole with mean orientation $\tilde{\phi}=\phi_c$ rotating within a wedge of half-angle $\tilde{\omega}$. (d)~The image produced by a pair of dipoles with separation angle $\delta$ is equivalent to that of a single dipole rotating within a wedge of half-angle $\tilde{\omega}$.}
    \label{fig:1}
\end{figure}

The absorption and emission of a fluorescent molecule can be modeled using a transition dipole moment \cite{novotny_2012_principles} 
$\bm{\mu} = \left(\mu_x,\mu_y,\mu_z\right) = \left( \sin{\theta}\cos{\phi}, \sin{\theta}\sin{\phi}, \cos{\theta} \right)$,
where $\theta$ and $\phi$ are polar and azimuthal angles, respectively, in spherical coordinates [Fig.~\ref{fig:1}(a)]. The $N$-pixel images 
\begin{equation}\label{eq:forwardModel_secM}
    \bm{I} = s\bm{B}\bm{m} + \bm{b}
\end{equation}
of the dipole produced by a microscope are a linear superposition \cite{backer_2014jpcb_extending} of its six basis images $\bm{B}$ with coefficients given by the second-order moments $\bm{m} = (m_{xx}, m_{yy}, m_{zz}, m_{xy}, m_{xz}, m_{yz} )$ of the transition dipole $\bm{\mu}$, where $s$ is the number of signal photons detected in the image plane, $m_{ij}=\left< \mu_i \mu_j \right>$ with $\{i,j\} \in \{x,y,z\}$, $\left< \cdot \right>$ denotes a temporal average over the acquisition time of the photodetector or camera, and $\bm{b} \in \mathbb{R}^{N}$ is the number of background photons in each pixel (see Supplemental Material \cite{Supplemental} and references therein \cite{tao_1969biopolymers_time-dependent,pieper_2011cpl_fluorescence,wang_2012accchemres_probing,stallinga_2015josaa_effect,backer_2015oe_determining,yang_2018nanolett_resolving,zhang_2019prl_fundamental,zhang_2021josaa_single-molecule,zhang_2022nanolett_resolving,volpato_2023nbt_extending,zhou_2024jpca_fundamental} for details). The basis images $\bm{B} \in \mathbb{R}^{N \times 6}$ only depend on the imaging system itself and can be calculated via vectorial diffraction theory \cite{backer_2014jpcb_extending, novotny_2012_principles, axelrod_2012jmicros_fluorescence}.

We consider two independent dipoles fixed in position and orientation with negligible separation in 3D space (the case of coupled dipoles has been explored elsewhere \cite{Hettich2002,Karuseichyk2024}). Without loss of generality, we assume their orientations lie in the $\mu_x$-$\mu_y$ plane with $\theta_1 = \theta_2 = 90^\circ$. The two in-plane dipoles can be parameterized by a centroid angle $\phi_c$ and separation angle $\delta$ in 2D, such that  $\bm{\mu}_1 = \left(\cos{(\phi_c-\delta/2)}, \sin{(\phi_c-\delta/2)},0 \right)$ and $\bm{\mu}_2 = \left( \cos{(\phi_c+\delta/2)}, \sin{(\phi_c+\delta/2)},0 \right)$. A simple rotation of the coordinate system can yield any arbitrarily oriented pair of dipoles in 3D \cite{Supplemental}.

If the two dipoles emit an equal number of photons, then the incoherent superposition of their images is given by $\bm{I} = \bm{I}_1 + \bm{I}_2 = s\bm{B}\left(\bm{m}_1 + \bm{m}_2\right)$. The second moments of this dipole pair are given by \cite{Supplemental} 
\begin{subequations}
    \label{eq:secM_dipolePair}
    \begin{align}
        m_{xx} &= \left[1 + \cos{\delta}\cos{(2\phi_c)}\right]/2, \\
        m_{yy} &= \left[1 - \cos{\delta}\cos{(2\phi_c)}\right]/2, \text{ and} \\
        m_{xy} &= \cos{\delta}\sin{(2\phi_c)}/2,  
    \end{align}
\end{subequations}
with $m_{zz}=m_{xz}=m_{yz}=0$. A microscope with x- and y-polarized imaging channels [Fig.~\ref{fig:1}(a)] would produce images of the pair shown in Fig.~\ref{fig:1}(b)(iii). In contrast, a single molecule with mean orientation $\tilde{\phi}$ wobbling uniformly within a wedge of half-angle $\tilde{\omega}$ within the $\mu_x$-$\mu_y$ plane would exhibit second moments given by  \cite{Supplemental}
\begin{subequations}
    \label{eq:secM_dipoleWobble}
    \begin{align}
        \left< \mu_x^2 \right> &= \left[ 1 + \sinc{(2\tilde{\omega})}\cos{(2\tilde{\phi})} \right]/2, \\
        \left< \mu_y^2 \right> &= \left[ 1 - \sinc{(2\tilde{\omega})}\cos{(2\tilde{\phi})} \right]/2, \text{ and} \\
        \left< \mu_x\mu_y \right> &= \sinc{(2\tilde{\omega})}\sin{(2\tilde{\phi})}/2,
    \end{align}
\end{subequations}
with $\sinc(x) = \sin{(x)}/x$ if $x \neq 0$ and zero otherwise.

Interestingly, we find that if $\phi_c = \tilde{\phi}$ and $\cos{\delta} = \sinc{2\tilde{\omega}}$, then the second moments of the dipole pair and the single wobbling dipole are identical. Figure~\ref{fig:1}(d) shows the one-to-one equivalence between the separation $\delta$ of the dipole pair and wobble angle $\tilde{\omega}$ of the SM. Critically, illuminating the sample with various excitation beam polarizations via ExM also cannot distinguish these cases \cite{Supplemental}. Thus, any instrument whose measurements are solely sensitive to second-order moments of a transition dipole is fundamentally unable to distinguish between a pair of dipoles from a single dipole, regardless of the centroid orientation $\phi_c$. The precise distribution of single-dipole wobble, e.g., uniform rotation or rotation within a harmonic potential \cite{lew_2013nanolett_rotational, stallinga_2015josaa_effect, chandler_2020josaa_spatio-angular} has no effect, as long as the second moments are indistinguishable. 

\begin{figure*}[thb!]
    \centering
    \includegraphics[width=6.4in]{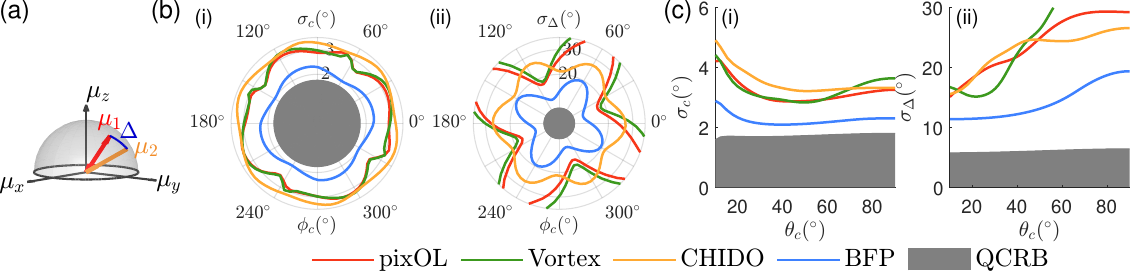}
    \caption{Precision of measuring the centroid orientation $\left(\theta_c, \phi_c\right)$ of and separation angle $\Delta$ between two fixed dipoles with 100 photons detected. (a)~Example dipole pair with $\bm{\mu}_1 = (\sin{(\theta_c-\theta_\Delta/2)} \cos{(\phi_c-\phi_\Delta/2)}$, $\sin{(\theta_c-\theta_\Delta/2)} \sin{(\phi_c-\phi_\Delta/2)}$, $\cos{(\theta_c-\theta_\Delta/2)} )$,  $\bm{\mu}_2 = (\sin{(\theta_c+\theta_\Delta/2)} \cos{(\phi_c+\phi_\Delta/2)}$,  $\sin{(\theta_c+\theta_\Delta/2)} \sin{(\phi_c+\phi_\Delta/2)}$,  $\cos{(\theta_c+\theta_\Delta/2)} )$, and separation $\Delta=\arccos{(\bm{\mu}_1^\intercal \bm{\mu}_2)}=20^\circ$. (b)~Best-possible precisions, quantified by the Cram\'{e}r-Rao bound (CRB) of measuring (i)~centroid orientation $\sigma_{c}$ [Eq.~(\ref{eq:sigma_c})] and (ii)~separation angle $\sigma_\Delta$ [Eq.~(\ref{eq:sigma_Delta})] for a dipole pair with $\theta_c = 60^\circ$. (c)~Best-possible precisions of measuring (i)~centroid orientation $\sigma_{c}$ and (ii)~separation angle $\sigma_\Delta$, averaged over all $\phi_c$. Red, pixOL DSF \cite{wu_2022optica_dipole-spread-function}; green, polarized vortex DSF \cite{ding_2021jpcb_single-molecule}; yellow, CHIDO \cite{curcio_2020nc_birefringent}; blue, back focal plane (BFP) imaging \cite{lieb_2004josab_single-molecule}. The gray regions show the precision limit bounded by the quantum CRB (QCRB).}
    \label{fig:2}
\end{figure*}

We next quantify how precisely the centroid orientation $\left(\theta_c, \phi_c\right)$ and separation angles $\left(\theta_\Delta, \phi_\Delta\right)$ of a dipole pair can be measured in 3D. For any unbiased estimator, the covariance of a set of estimates is bounded by the classical and quantum Cram\'{e}r-Rao bounds (CRBs) \cite{kay_1993_fundamentals, helstrom_1976_quantum, tsang_2016prx_quantum, liu_2020jpamt_quantum}, given by
\begin{equation}
    \mathrm{Cov}(\bm{\hat{\Theta}}) \succeq \bm{\mathcal{J}}^\mathrm{-1} \succeq \bm{\mathcal{K}}^{-1}, 
\end{equation}
where $\bm{\hat{\Theta}} = \left(\hat{\theta}_c, \hat{\phi}_c, \hat{\theta}_\Delta, \hat{\phi}_\Delta\right)$ represents the set of unbiased estimates, $\bm{\mathcal{J}}$ and $\bm{\mathcal{K}}$ represent the classical and quantum Fisher information (FI) matrices, respectively, and $\succeq$ denotes a generalized inequality such that $\left( \mathrm{Cov}(\bm{\hat{\Theta}}) - \bm{\mathcal{J}}^\mathrm{-1} \right)$ and $\left( \bm{\mathcal{J}}^\mathrm{-1} - \bm{\mathcal{K}}^{-1} \right)$ are both positive semidefinite. 
We calculate the precision $\sigma_c$ of measuring the centroid orientation by computing the average standardized generalized variance (SGV) of the CRB and converting it to an equivalent angle on the orientation sphere [Fig.~\ref{fig:1}(a)], as \cite{chandler_2017oe_single-fluorophore, wu_2022optica_dipole-spread-function}
\begin{equation}\label{eq:sigma_c}
    \sigma_{c} = 2 \arcsin\left[ \sqrt{ \frac{\sin{\theta_c} \left(\det \left\{ \bm{\mathcal{J}}^{-1} \right\}_{1:2,1:2} \right)^{1/2}}{4\pi} } \right].
\end{equation}
Similarly, we represent the precision $\sigma_\Delta$ of measuring the 3D separation angle $\Delta = \arccos{(\bm{\mu}_1^\intercal \bm{\mu}_2)}$ via
\begin{equation}\label{eq:sigma_Delta}
    \sigma_\Delta = \sqrt{\bm{J}^\intercal \bm{\mathcal{J}}^{-1} \bm{J}},
\end{equation}
with the Jacobian
\begin{equation}\label{eq:jacobian_separation_simplified}
    \bm{J} = \left(\partial \Delta / \partial \theta_c, \partial \Delta / \partial \phi_c, \partial \Delta / \partial \theta_\Delta, \partial \Delta / \partial \phi_\Delta \right) \in \mathbb{R}^{4}.
\end{equation}
We also use these transformations to compute analogous best-possible precisions for any imaging system using quantum FI (QFI) $\bm{\mathcal{K}}$. 
Note that classical FI $\bm{\mathcal{J}}$ is calculated for specific microscopes imaging molecules at specific orientations $\bm{\Theta}$ \cite{chao_2016josaa_fisher}, and QFI provides a universal precision bound for any imaging system \cite{helstrom_1976_quantum, lupo_2016prl_ultimate}. While one cannot guarantee that a method can be devised to reach the QFI limit, comparing the CRB achieved by existing techniques to the quantum CRB (QCRB) \cite{Supplemental} provides a useful context for evaluating performance.

We find that imaging the back focal plane (BFP) \cite{lieb_2004josab_single-molecule} exhibits the best precision for measuring the centroid orientation of a dipole pair and is closest to the precision limit given by QCRB, in accordance with previous observations \cite{zhang_2020prr_quantum}. Other widely-used, state-of-the-art DSFs, including CHIDO \cite{curcio_2020nc_birefringent}, vortex \cite{ding_2021jpcb_single-molecule, hulleman_2021nc_simultaneous} and pixOL \cite{wu_2022optica_dipole-spread-function} all show excellent and uniform performance in estimating mean orientation [Fig.~\ref{fig:2}(a)(i) and (b)(i)], with precisions within \textasciitilde{}60\% of the QCRB limit. However, all techniques perform poorly in estimating the separation angle between the dipole pair [Fig.~\ref{fig:2}(a)(ii) and (b)(ii)], giving precisions of 20-40$^\circ$ with 100 photons detected.

Given these observations, two questions naturally arise: Is it possible to design an imaging system that can distinguish a pair of molecules from a single molecule? And, once a dipole pair has been identified, how precisely can this system measure their angular separation? In response, we propose a natural extension of ExM and DSF engineering: using them jointly for orientation imaging. Our analysis shows that the images $\bm{I}$ collected by such a system are linear with respect to fourth-order moments $\bm{q}\in\mathbb{R}^{15}$ of the transition dipole $\bm{\mu}$ (see Supplemental Material \cite{Supplemental} for a derivation \cite{stallinga_2015josaa_effect, munger_2023oc_influence}). That is, if $L$ polarization states illuminate the sample sequentially and $N$-pixel images are collected for each DSF, then 
\begin{equation}
    \bm{I}  = a\bm{H}\bm{q} + \bm{b},
\end{equation}
where $a$ incorporates the absorption cross-section and quantum yield of the molecule, the wavelength of the excitation beam, etc.\ \cite{Supplemental,moerner_2003rsi_methods}. The forward operator $\bm{H} \in \mathbb{R}^{NL\times 15}$ represents the imaging system's response to each fourth moment $q$ and incorporates the impacts of excitation beam polarization and intensity and DSF engineering on the final images.

\begin{figure}[b]
    \centering
    \includegraphics[width=8.6cm]{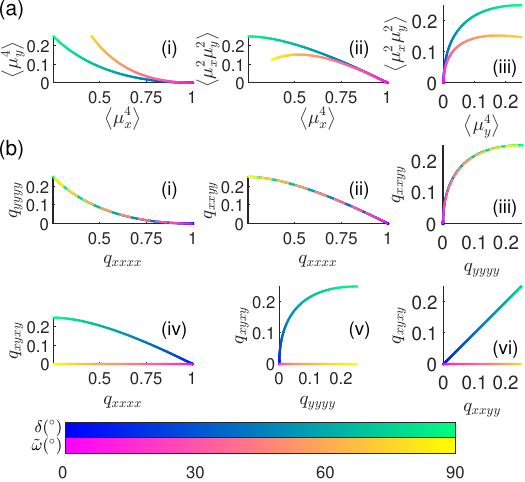}
    \caption{Fourth moments of a fixed dipole pair with separation $\delta$ (blue-green curve) compared to a single dipole wobbling within a wedge of half-angle $\tilde{\omega}$ [pink-yellow curve, Fig.~\ref{fig:1}(c)] exhibiting (a) slow rotational diffusion (i.e., long rotational correlation time relative to the excited state lifetime) and (b) fast rotational diffusion. All dipoles are oriented within the $\mu_x$-$\mu_y$ plane. (a)~Fourth moments of a dipole pair and a single slowly wobbling dipole shown as 2D correlations between (i)~$\left<\mu_x^4\right>$ vs.\ $\left<\mu_y^4\right>$, (ii)~$\left<\mu_x^4\right>$ vs.\ $\left<\mu_x^2\mu_y^2\right>$, and (iii)~$\left<\mu_y^4\right>$ vs.\ $\left<\mu_x^2\mu_y^2\right>$. (b)~Fourth moments $q_{ijkl}$ of a dipole pair and a single quicly wobbling dipole are shown as 2D correlations between (i)~$q_{xxxx}$ vs.\ $q_{yyyy}$, (ii)~$q_{xxxx}$ vs.\ $q_{xxyy}$, (iii)~$q_{yyyy}$ vs.\ $q_{xxyy}$, (iv)~$q_{xxxx}$ vs.\ $q_{xyxy}$, (v)~$q_{yyyy}$ vs.\ $q_{xyxy}$, and (vi)~$q_{xxyy}$ vs.\ $q_{xyxy}$.}
    \label{fig:3}
\end{figure}
We now augment our previous dipole pair analysis and assume that they lie in the $\mu_x$-$\mu_y$ plane with a centroid orientation pointing along the $\mu_x$ axis. We therefore have $\bm{\mu}_1 = (\cos{(-\delta/2)}$, $\sin{(-\delta/2)}$, $0)$ and $\bm{\mu}_2 = (\cos{(\delta/2)}$, $\sin{(\delta/2)}$, $0)$, and again, simple coordinate rotations applied to this pair can produce any arbitrary configuration. The photons collected from this pair still sum incoherently in the image plane, such that $\bm{I} = \bm{I}_1 + \bm{I}_2 = a\bm{H}\left(\bm{q}_1 + \bm{q}_2\right)$. Due to symmetry, there are only three non-zero elements of $\bm{q}$, given by $\mu_x^4 = \cos^4\left( \delta/2 \right)$, $\mu_y^4 = \sin^4\left( \delta/2 \right)$, and $\mu_x^2\mu_y^2 = \sin^2\left( \delta \right)/4$ \cite{Supplemental}.
For a single wobbling molecule with mean orientation $\tilde{\bm{\mu}} = (1,0,0)$ and wobble angle $\tilde{\omega}$ in the $\mu_x$-$\mu_y$ plane, we analyze two limiting cases. When a molecule's rotational correlation time is long relative to the excited state lifetime, there are only 3 non-trivial fourth moments $\left<\mu_x^4\right>$, $\left<\mu_y^4\right>$ and $\left<\mu_x^2\mu_y^2\right>$.
In the opposite limit of fast diffusion, the absorption and emission transition dipoles are decoupled, such that $q_{ijkl} = \left<\mu_{a,i}\mu_{a,j}\right>\left<\mu_{e,k}\mu_{e,l}\right>$, where $\mu_{a,i}$ and $\mu_{e,k}$ represent the $i^\text{th}$ and $k^\mathrm{th}$ components of the absorption and emission transition dipole, respectively, with $ i,j,k,l \in \{x,y,z\}$. In this case, there are 4 non-trivial fourth moments $q_{xxxx}$, $q_{yyyy}$, $q_{xxyy}$ and $q_{xyxy}$ \cite{Supplemental}.

Importantly, the fourth moments $\tilde{\bm{q}}$ of the SM are functionally distinct from those of a dipole pair, regardless of the speed of the rotational diffusion. In fact, their only intersection occurs when $\delta=\tilde{\omega}=0$ (Figure~\ref{fig:3}); naturally, a pair of dipoles with identical orientations is equivalent to a single fixed dipole. For any $\delta \neq 0$, the fourth moments of any dipole pair are \emph{unique} from those of a single wobbling molecule. 
Therefore, by measuring fourth moments with sufficient precision, one may distinguish dipole pairs from single molecules even if they spatially coincide.

\begin{figure}[bth]
    \centering
    \includegraphics[width=8.6cm]{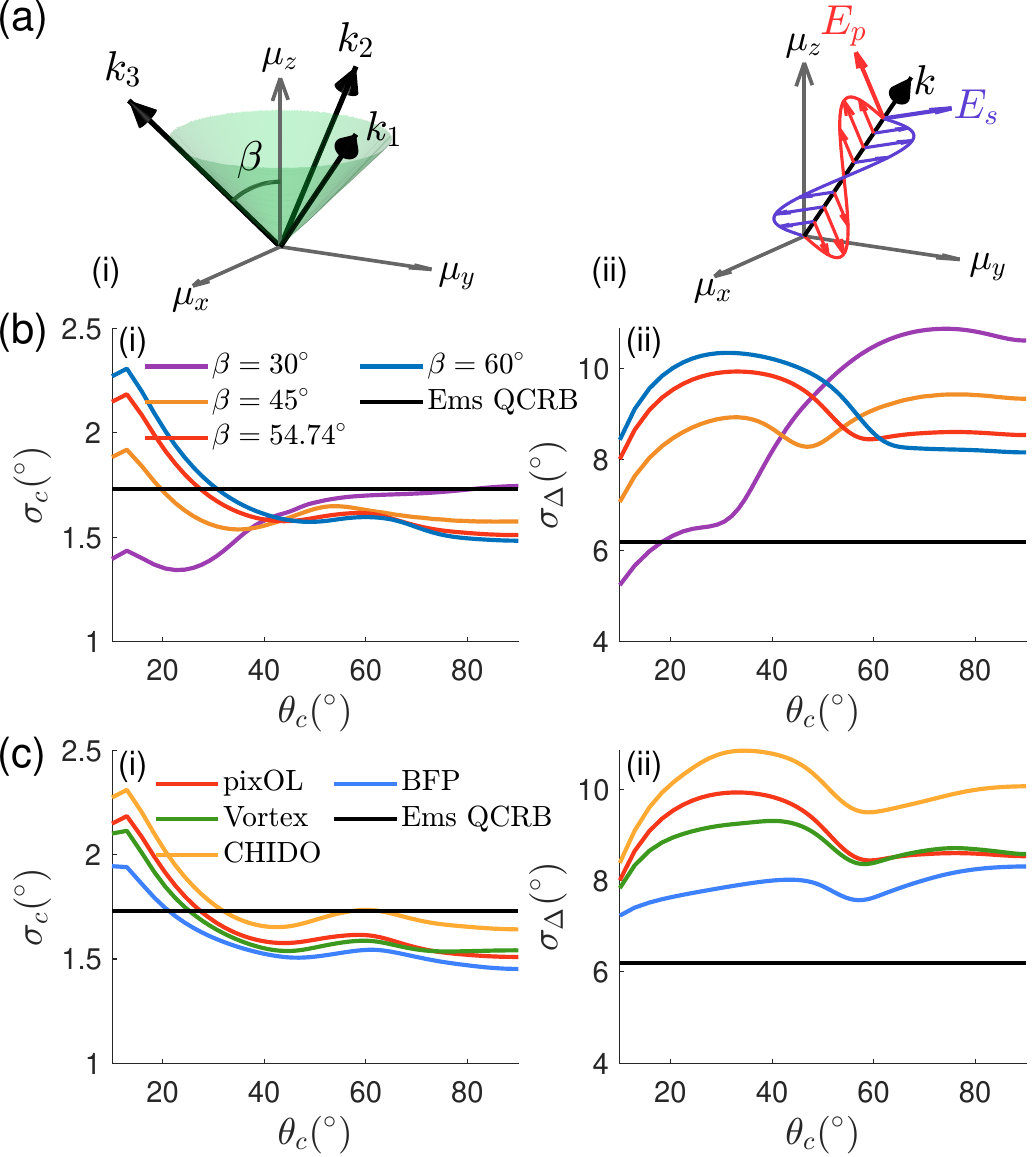}
    \caption{Performance of using combined excitation modulation (ExM) and dipole-spread function (DSF) engineering. (a)~Schematic of (i) 3 designed wavevectors $\bm{k}$ and (ii) their associated s- and p-polarized states for ExM. The three $\bm{k}$ vectors lie on a cone oriented along the $\mu_z$ direction with half-angle $\beta$, and the relative azimuthal angle between any pair is $120^\circ$. (b)~Estimation precision for the (i) centroid orientation of and (ii) separation between a pair of dipoles using ExM-pixOL with various $\beta$ angles. Purple, $\beta = 30^\circ$; orange, $\beta = 45^\circ$; red, $\beta = \arccos{(1/\sqrt{3})} \approx 54.74^\circ$; blue, $\beta = 60^\circ$. (c)~Same as (b) for different DSFs and $\beta = \arccos{(1/\sqrt{3})} \approx 54.74^\circ$. Red, pixOL DSF \cite{wu_2022optica_dipole-spread-function}; green, polarized vortex DSF \cite{ding_2021jpcb_single-molecule}; yellow, CHIDO \cite{curcio_2020nc_birefringent}; blue, back focal plane (BFP) imaging \cite{lieb_2004josab_single-molecule}. The black line in (b) and (c) refers to the QCRB of fluorescence emission (ems, i.e., DSF engineering, Fig.~\ref{fig:2}). Precisions are calculated for a dipole pair with separation $\Delta=20^\circ$ and are averaged over $\phi_c \in [0^\circ, 360^\circ]$ with 100 photons detected and zero background.}
    \label{fig:4}
\end{figure}

We now quantify the performance of measuring the orientations of and separation between dipole pairs when combining ExM with various engineered DSFs. For ExM, we parameterize the sequence of illumination polarizations using three equally spaced wavevectors $\bm{k}_1$, $\bm{k}_2$, and $\bm{k}_3$ lying on the surface of a cone with half-angle $\beta$ oriented along the $\mu_z$ axis [Fig.~\ref{fig:4}(a)(i)]. Each wavevector is associated with s- and p-polarized light [Fig.~\ref{fig:4}(a)(ii)], thereby resulting in 6 pumping polarizations. The number of photons absorbed by the molecule and therefore detected by a camera is related to both the pumping electric field $\bm{E}_i$ and the dipole's orientation $\bm{\mu}$ as $s \propto \sum_{i=1}^6 \left| \bm{E}_i^\intercal \bm{\mu} \right|^2$ \cite{Supplemental}.

With 100 signal photons and zero background detected in total across all measurements, ExM-pixOL achieves \textasciitilde{}1.8-degree precision (a 50\% improvement) for measuring centroid orientation and remains stable across the orientation hemisphere [Fig.~\ref{fig:4}(b)]. For measuring dipole separation, combining ExM with pixOL leads to a 2- to 4-fold precision improvement. Among all excitation schemes, using a ``magic'' illumination angle $\beta = \arccos{(1/\sqrt{3})} \approx 54.74^\circ$ enables approximately uniform measurement performance for any centroid polar angle $\theta_c$, because this $\beta$ angle excites fluorophores evenly across all possible orientations \cite{Supplemental}. 
In addition, various DSFs exhibit similar performance for $\beta = \arccos{(1/\sqrt{3})}$, [Fig.~\ref{fig:4}(c)]. Importantly, combining ExM and DSF engineering can achieve performance superior to the best-possible engineered DSF, according to the QCRB. Interestingly, ExM-BFP imaging achieves \textasciitilde{}13\% better precision on average than ExM-DSF engineering techniques, suggesting that it can improve angular resolution at the cost of requiring spatial point scanning for imaging \cite{thiele_2020acsnano_confocal}.

In summary, we analyzed the fundamental physical limits of measuring the orientations of a pair of coinciding dipole emitters using polarization-sensitive single-molecule imaging. Since the second moments of a pair of dipole emitters and a rotating molecule are identical, neither excitation modulation nor DSF engineering techniques can distinguish between the two scenarios without prior knowledge. However, by increasing the dimensionality of the measurement to the fourth-moment space via combined ExM and DSF engineering, pairs of dipoles produce camera images that are unique from those of a single wobbling molecule, thereby solving the degeneracies that hamper existing techniques. We further remark that time-resolved fluorescence anisotropy \cite{lakowicz_principles_2006, jameson_2010chemrev_fluorescence} may be viewed as a fourth-moment measurement, where both polarized illumination and detection are required to measure rotational dynamics and distinguish a pair of dipoles from a single wobbling molecule \cite{Supplemental}. We demonstrated that combined ExM-DSF engineering exhibits superior performance in measuring the centroid orientation of and separation between a pair of molecules.  In general, the optimal combination of ExM and engineered DSF depends on the sample of interest, such as its thickness, the anticipated distribution of emitter orientations, whether 2D or 3D spatial information is desired, and fluorophore brightness. As more parameters are measured, e.g., going from 2D orientations in the xy plane to orientations in full 3D, performance tradeoffs are often necessary \cite{Zhang-single-molecule-2021b-Comparison,zhang_single-molecule_2024}.
Our work lays the foundation for developing optimal techniques for resolving spatially coinciding dipole emitters and precisely measuring their rotational dynamics in the crowded environments typical of biological samples. 

\begin{acknowledgments}
We thank Tingting Wu, Weiyan Zhou, and Suraj Deepak Khochare for insightful discussions. This work was supported by the National Institute of General Medical Sciences of the National Institutes of Health under Award Number R35GM124858.
\end{acknowledgments}


\bibliography{bibliography}



\end{document}